\documentclass[12pt]{article}
\usepackage{epsf,amsfonts,amssymb,epsfig,amsmath}
\renewcommand{\text}[1]{#1}

\newcommand{\be}{\begin{equation}}
\newcommand{\ee}{\end{equation}}
\newcommand{\ben}{\begin{displaymath}}
\newcommand{\een}{\end{displaymath}}
\newcommand{\bea}{\begin{eqnarray}}
\newcommand{\eea}{\end{eqnarray}}
\newcommand{\bean}{\begin{eqnarray*}}
\newcommand{\eean}{\end{eqnarray*}}
\newcommand{\nn}{\nonumber \\}
\newcommand{\ba}{\begin{array}}
\newcommand{\ea}{\end{array}}
\newcommand{\bi}{\begin{itemize}}
\newcommand{\ei}{\end{itemize}}

\newcommand{\reef}[1]{(\ref{#1})}

\def\G{\Gamma}

\def\e{\epsilon}

\def\otaula{\begin{tabular}}
\def\ctaula{\end{tabular}}

\def\bnum{\begin{enumerate}}
\def\enum{\end{enumerate}}

\def\CR{\mathcal{R}}
\def\CM{\mathcal{M}}

\def\8M{$\CM_8$}

\def\be{\begin{equation}}
\def\ee{\end{equation}}
\def\G{\Gamma}

\def\ei{e^{\underline{i}}}

\def\e1{e^{\underline{1}}}
\def\1u{\underline{1}}
\def\2u{\underline{2}}

\def\0u{\underline{0}}
\def\e{\epsilon}
\def\target{$\CR^{1,1}\times \mathcal{M}_8$ }
\def\target2{$\CR^{1,1}\times \mathcal{M}_8$,}
\def\9G{\G_{\underline{9}}}

\newcommand{\bbZ}{{\mathbb{Z}}}

\def\1f{f_1^{1/2}}
\def\2f{f_2^{1/2}}
\def\4f{f_4^{1/2}}

\begin{document}

\makeatletter
\renewcommand{\theequation}{\thesection.\arabic{equation}}
\@addtoreset{equation}{section}
\makeatother

\baselineskip 18pt

\begin{titlepage}

\vfill

\begin{flushright}
Imperial/TP/2006/JG/04\\
hep-th/0611219\\
\end{flushright}

\vfill

\begin{center}
   \baselineskip=16pt
   {\Large\bf  Properties of some conformal field theories with M-theory duals}
   \vskip 2cm
      Jerome P. Gauntlett$^{1,2}$, Eoin \'{O} Colg\'{a}in$^1$
        and Oscar Varela$^3$
         \vskip .6cm
      \begin{small}
      $^1$\textit{Theoretical Physics Group, Blackett Laboratory, \\
        Imperial College, London SW7 2AZ, U.K.}
        \end{small}\\*[.6cm]
      \begin{small}
      $^2$\textit{The Institute for Mathematical Sciences, \\
        Imperial College, London SW7 2PE, U.K.}
        \end{small}\\*[.6cm]
      \begin{small}
      $^3$\textit{Departamento de F\'\i sica Te\'orica and IFIC,\\
            Universidad de Valencia-CSIC, 46100-Burjassot (Valencia), Spain}
        \end{small}
   \end{center}

\vfill

\begin{center}
\textbf{Abstract}
\end{center}

\begin{quote}
By studying classes of supersymmetric solutions of
$D=11$ supergravity with $AdS_5$ factors, we determine some properties
of the dual four-dimensional $N=1$ superconformal field theories.
For some explicit solutions we calculate the central charges and also the conformal
dimensions of certain chiral primary  operators arising
from wrapped membranes. For the most general class of solutions
we show that there is a consistent Kaluza-Klein truncation to minimal
$D=5$ gauged supergravity. This latter result allows
us to study some aspects of the dual strongly coupled thermal plasma with a non-zero
$R$-charge chemical potential and, in particular, we show that the ratio of
the shear viscosity to the entropy density has the universal value of $1/4\pi$.
\end{quote}

\vfill

\end{titlepage}
\setcounter{equation}{0}


\section{Introduction}

A classification of the most general supersymmetric solutions of $D=11$ supergravity
that consist of a warped product of $AdS_5$ with a six-dimensional compact manifold $M_6$
was presented in \cite{gmsw1}. Via the AdS/CFT correspondence these correspond to
the most general class of four-dimensional conformal field theories (CFTs) with $N=1$ supersymmetry
that have M-theory duals.
It was shown that $M_6$ has a canonical $SU(2)$ structure that is related to
the four-form flux and the warp factor, via some differential conditions. In the special sub-class
where an almost complex structure on $M_6$ is assumed to be integrable,
the differential conditions were integrated in \cite{gmsw1} and a rich class of
explicit solutions were constructed. For these explicit solutions the manifold $M_6$ is
topologically an $S^2$ bundle over a four-dimensional base manifold $B_4$.

For the special case of explicit solutions where $B_4=S^2\times T^2$,
after dimensional reduction and T-duality one
obtains the $AdS_5\times Y^{p,q}$ solutions of type IIB string theory where
$Y^{p,q}$ are an infinite class of Sasaki-Einstein metrics \cite{gmsw1,gmsw2}.
For this class of solutions the dual conformal field theories were
identified in \cite{SEduals} and this has led to many further developments.
It would be desirable to have a similar understanding of the conformal
field theories dual to the other explicit $AdS_5\times_w M_6$ solutions discovered
in \cite{gmsw1}. Here, as a step in this direction, one of our goals
will be to study the explicit solutions of \cite{gmsw1} in more detail and extract some
quantities of the dual CFT. We begin by determining the conditions imposed
by quantisation of the four-form flux and then use this to calculate
the central charges of the CFT. For the $B_4=S^2\times T^2$ case,
we recover the known formula \cite{gmsw2} for the central charge which has, for fixed $p$ and $q$,
an $N^2$ scaling corresponding to the dual
CFT being a gauge theory. For all other $B_4$ we find that the central charge
scales like $N^3$, just as for the six-dimensional $(2,0)$ field theory that lives
on M5-branes \cite{Klebanov:1996un}, indicating that the dual CFT is something exotic.

We next study supersymmetric probe membranes that wrap two-cycles in $M_6$,
which correspond to chiral primary operators in the dual CFT.
We first identify a generalised calibration \cite{gpt} that can be constructed
from the Killing spinors. We argue that if the membrane
worldvolume is calibrated by the generalised calibration then it is supersymmetric.
We determine the supersymmetric cycles for the explicit solutions and calculate
the conformal dimensions of the chiral primaries finding that they all scale like $N$.

The final topic is a study of some aspects of the strongly coupled thermal plasma of the CFTs.
Recall that the long distance, low-frequency behaviour of an interacting quantum field
theory at finite temperature is expected to be well-described by fluid dynamics.
Starting with the work of \cite{Policastro:2001yc} there has been a number of
papers that calculate various transport co-efficients, such as shear viscosity, $\eta$, and
diffusion constants, for quantum field theories that have gravity duals.
This is achieved by studying fluctuations about
black hole geometries and employing the AdS/CFT correspondence.
Following \cite{Policastro:2001yc,Policastro:2002se,Herzog:2002fn,Kovtun:2003wp}
it was shown that, in the absence of chemical potentials,
the ratio of shear viscosity to entropy density, $\eta/s$, is universal and equal to $1/4\pi$
\cite{Buchel:2003tz,Kovtun:2004de}.
For the case of $N=4$ super-Yang-Mills theory it
was subsequently shown that this universal result persists in the presence of non-zero $R$-charge chemical
potentials in \cite{Mas:2006dy,Son:2006em,Maeda:2006by} (a calculation for the theory
living on M2-branes was performed in \cite{Saremi:2006ep}).
Most recently, it was shown in \cite{Buchel:2006gb} that $\eta/s=1/4\pi$ is also valid
for the $N=1$ gauge theories that are dual to $AdS_5\times SE_5$ solutions of type IIB supergravity,
where $SE_5$ is an arbitrary five-dimensional Sasaki-Einstein space.

Here we will show that for the CFTs dual to the most general supersymmetric $AdS_5$ solutions of \cite{gmsw1}
(i.e. not just the explicit solutions) we also have $\eta/s=1/4\pi$.
This is noteworthy since, as remarked above, the CFTs
dual to these solutions must, in general, be quite different to the gauge theories that
are dual to the $Y^{p,q}$ spaces. In addition we show that the speed of sound in the plasma is
$1/{\sqrt 3}$.

Our approach for obtaining these results will follow that of \cite{Buchel:2006gb}.
It was shown in \cite{Buchel:2006gb} that there is a consistent Kaluza-Klein
truncation of type IIB supergravity on any $D=5$ Sasaki-Einstein space to minimal
five-dimensional gauged supergravity. The gauge field of the gauged supergravity corresponds
to the $R$-symmetry of the CFT. This means that any solution of the gauged supergravity
will give rise to an exact solution of type IIB supergravity. This is remarkable since
such consistent truncations are not common (see \cite{pope} for a relevant discussion).
By studying some charged black hole solutions of the five-dimensional gauged supergravity,
the authors of \cite{Buchel:2006gb} then used the consistent truncation to extract
results about the hydrodynamics of the thermal plasma of the dual CFT.

Here we will show that for the most general $AdS_5\times_w M_6$ solutions of \cite{gmsw1} there
is also a consistent truncation to five-dimensional minimal gauged
supergravity.
This result allows us to directly import some of the analysis of \cite{Buchel:2006gb} to
obtain the properties about the thermal plasma mentioned above.

The plan of the rest of the paper is as follows. Section 2 analyses flux quantisation
and derives the central charges for some of the explicit solutions of \cite{gmsw1}.
Section 3 discusses supersymmetric wrapped membranes and, for the
explicit solutions of section 2, we
calculate the conformal dimension of the associated chiral primary
operators.
Section 4 discusses the consistent Kaluza-Klein truncation and this is used
to normalise the $R$-charge of the wrapped membranes and also to
study the hydrodynamics of the dual CFT. Section 5 briefly concludes.

\section{Flux quantisation and Central charges}

The class of supersymmetric solutions of $D=11$ supergravity that we shall consider in this paper
were first analysed in \cite{gmsw1}. The $D=11$ metric is a warped product
\begin{equation}
\label{warpedansatz}
ds^{2} = L^2e^{2 \lambda}[ds^{2}(AdS_{5}) + ds^{2}(M_{6})],
\end{equation}
where $\lambda$ only depends on the coordinates of $M_6$.
The metric $ds^2(AdS_5)$ is that of a unit radius $AdS_5$ and the
length scale $L$ fixes the overall scale.
The four-form flux $G_{(4)}$ is a four-form just on $M_6$.
In general, it was shown that
$M_6$ has an $SU(2)$ structure constructed from the Killing spinors
that is specified by two one-forms $K^1$, $K^2$, a real two-form $J$,
a complex two-form $\Omega$ and a scalar $\cos\zeta$. The metric on $M_6$ can
be written
\be\label{anexc1}
ds^2(M_6)=e^ie^i +(K^1)^2+(K^2)^2 ,
\ee
with $J=e^{1}e^{2}+e^{3}e^{4}$ and $\Omega=(e^1+ie^2)(e^3+ie^4)$ where the products of forms
are taken to be wedge products. The vector dual to
$\cos\zeta K^2$ is a Killing vector which is related to the $R$-symmetry of
the dual CFT as we shall discuss later. Introducing coordinates where this Killing vector
is given by $3\partial_\psi$ one can show that the
metric can be written
\be\label{anexc}
ds^2(M_6)=g^4_{ij}dx^idx^j+e^{-6\lambda}\sec^2\zeta dy^2+\frac{1}{9}\cos^2\zeta(d\psi+\rho)^2 ,
\ee
where $g^4_{ij}$, $\lambda$, $\zeta$ and $\rho$ are all functions of $x^i$ and $y$.
In these coordinates we have $K^1=e^{-3\lambda}\sec\zeta dy$ and $K^2=(1/3)\cos\zeta (d\psi+\rho)$.
The expression for the most general four-form flux can be found in section 4.
The precise conditions that need to be imposed
in order to get a supersymmetric solution,
which include $2y=e^{3\lambda}\sin\zeta$, can be found in \cite{gmsw1}.

Some of this paper will focus on the explicit solutions of \cite{gmsw1}
which were constructed by demanding that the natural almost complex structure
on $M_6$ is integrable and hence $M_6$ is a complex manifold. In this case, topologically,
$M_6$ is an $S^2$ bundle over a four-dimensional base manifold $B_4$ that is either
K\"ahler-Einstein with positive curvature, and hence $CP^2$,
$S^2\times S^2$ or a del-Pezzo $dP_k$, $k=3,\dots ,8$, or alternatively
a product space $S^2\times S^2$, $S^2\times T^2$ or $S^2\times H^2$ (one can also replace $H^2$ with $H^2/\Gamma$).
The $S^2$ fibration can be obtained by taking the canonical line-bundle
of the four-dimensional K\"ahler base space and adding a ``point at infinity''
to each of the fibres. The detailed form of the metric on $M_6$ and the four-form for
the explicit solutions of \cite{gmsw1} that we shall study here will be given shortly.

In order to get a
good solution of $M$-theory we need to ensure that the four-form flux is properly
quantised. In particular we demand that
\begin{equation}
\label{M5charge}
N_{\Sigma_4}\equiv \frac{1}{(2 \pi)^{3} l_{11}^{3}} \int_{\Sigma_{4}} G_{(4)} \in \mathbb{Z}
\end{equation}
for any four-cycle $\Sigma_{4}$ on $M_6$, where $l_{11}$ is the $D=11$
Planck length. As usual, this leads to a quantisation condition on
the length scale $L$.

The central charge of the dual conformal field theory can be obtained from the formula \cite{weyl}
\begin{equation}
\label{ccharge}
c= \frac{\pi R_{AdS_5}^{3}}{8 G_{5}} .
\end{equation}
where $G_5$ is the effective five-dimensional Newton constant and
$R_{AdS_5}$ is the radius of $AdS_5$ in the five-dimensional
theory. Taking the warp factor into account, a short calculation
shows that \be\label{ccharge2}
c=\frac{1}{2^7\pi^6}\left(\frac{L}{l_{11}}\right)^9\int_{M_6}
d^{6}x \sqrt{g(M_6)}e^{9 \lambda} . \ee

We now consider three explicit examples in turn. When $B_4$ is K\"ahler-Einstein, when $B_4 =S^2\times S^2$
and when $B_4=T^2\times S^2$. The case $B_4=S^2\times H^2$ is very similar to the $S^2\times S^2$ case
and for simplicity of presentation we omit the details.

\subsection{$\mathbf{B_4=KE_4}$}

For this case
the general solutions of \cite{gmsw1} depend on two parameters $b$ and $c$.
The internal metric is given by
\begin{equation}
\label{KEmetric}
ds^{2}(M_6) = \frac{e^{-6\lambda}}{3} (b-y^{2}) ds^{2} (KE_{4}) + e^{-6
\lambda }\sec^{2} \zeta d y^{2} + \frac{1}{9 } \cos^{2}
\zeta D\psi^{2},
\end{equation}
with
\begin{eqnarray}
e^{ 6 \lambda } &=& \frac{2(b - y^{2})^{2}}{cy+2b + 2y^{2}}, \\
\label{eqn}
\cos^{2} \zeta &=& \frac{-3y^{4} - 2cy^3-6b y^{2} + b^2}{(b-y^{2})^{2}}
\end{eqnarray}
and $ds^2(KE_4)$ is (a $y$ independent) four-dimensional K\"ahler-Einstein metric
with positive curvature.
We shall normalise this metric so
that $\mathcal{R} = J_{KE}$ where $\mathcal{R}$ is the Ricci-form and $J_{KE}$ is the
K\"{a}hler form. We also have $D\psi\equiv d\psi+P$ where $P$ is
the connection on the canonical bundle of the $KE_4$, i.e. $dP={\cal R}$.
The $S^2$ fibre is parametrised by $y$ and $\psi$: the coordinate $\psi$ is periodic
with period $2\pi$ and $y$ lies in an interval $y_1\le y\le y_2$ where
the $y_i$ are appropriate roots of the quartic in the numerator of $\cos^{2} \zeta$.
The four-form flux is given by
\begin{equation}\label{eq4f}
\frac{1}{L^3}G_{(4)} = \frac{4 y^{3} + 3cy^2+12b y+bc}{18 (y^{2} - b)} {vol}_{KE_{4}} +
\frac{y^{4}-6b y^{2} -2bcy- 3b^2}{9 (y^{2}-b)^{2}} {J_{KE}}
dy D\psi ,
\end{equation}
where the products of forms are taken to be wedge products.

In order to keep the analysis simple, we will now restrict to the $c=0$ class of
solutions. For this class one can rescale to set $b=1$.
The range of $y$ is now easily determined
from
$3 y^4 + 6 y^2 - 1 = 0$:
we take $y_1\le y\le y_2$ with
\begin{equation}
\label{KEroots}
y_{1,2} = \mp\left[  \frac{2}{\sqrt{3}}-1 \right]^{1/2}.
\end{equation}

To implement the flux quantisation condition \reef{M5charge} we need to identify a basis of
four-cycles. If we let the two-cycles $\Sigma_{a}$ be a basis for $H_{2}(KE_{4},\mathbb{Z})$,
then we can take the basis to be $C_a, C_N$ where $C_a$ are the four-cycles obtained
by considering the $S^2$ fibration over the two-cycles $\Sigma_a$ and $C_N$ is the
four-cycle obtained by restricting $y$ to lie at the ``north pole'' of the two-sphere
fibre by setting $y=y_2$. Instead of $C_N$ we could
also consider the four-cycle $C_S$
sitting at the ``south pole'' of the fibre by setting $y=y_1$.

It will be useful to introduce some notation concerning the $KE_4$ base space.
In particular we have
\begin{equation}
n (\Sigma_{a})  =
\frac{1}{2 \pi}
\int_{\Sigma_{a}} \mathcal{R}=mn_a ,
\end{equation}
where the positive integer $m$ is known as the Fano index
of $KE_4$ and is the largest positive integer such that all of the $n_a$ are integers.
We also define
\begin{equation}
M = \frac{1}{4\pi^2}\int_{KE_{4}} \mathcal{R}\wedge \mathcal{R}
\end{equation}
and we note that $M$ is always divisible by $m^2$.
For further discussion see e.g. appendix B of \cite{gmmw3}, but
we note here that the explicit values for $(m,M)$
for $CP^2$ are $(3,9)$, for $S^2\times S^2$ are $(2,8)$ and
for $dP_k$, $k=3,\dots, 8$ are $(1,9-k)$.

We now calculate the flux threading through the various cycles.
Using (\ref{M5charge}), we find
\begin{eqnarray}
\label{north}
N_{C_N}
&=&  -\left(\frac{y_{2}L^{3}(2+\sqrt{3})}{18 \pi l_{11}^{3} }
\right) M,\nn
N_{C_{a}}
&=& -\left(\frac{y_{2}L^{3}(2+\sqrt{3})      }{18 \pi l_{11}^{3}}
\right)
 2mn_{a}.
\end{eqnarray}
We note that $N_{C_S}=-N_{C_N}$. In order to ensure that $N_{C_N}$
and $N_{C_a}$ are indeed integers, we choose the length scale $L$ to
be given by
\begin{equation}
\label{tunedm}
\frac{y_{2}L^{3}(2+\sqrt{3}) }{18 \pi l^{3}_{11}}  = \frac{N}{h},
\end{equation}
where $N$ is an arbitrary integer and $h$ is the highest common factor
of $M$ and $2m$. We then have $N_{C_N}=-(M/h)N$ and $N_{C_a}=-(2m/h)n_aN$.

Using (\ref{ccharge2}),
we are now in a position to determine the
central charge in terms of these brane charges and we find
\begin{equation}
\label{c1}
c = 9 (3 \sqrt{3}-5)\frac{M}{h^3} N^{3}.
\end{equation}

\subsection{$\mathbf{B_4=S^{2} \times S^{2}}$:}
When $B_4=S^2\times S^2$ the general solutions of \cite{gmsw1}
depend on three parameters $a_1, a_2$ and $c$. The metric takes the form
\begin{equation}
\label{S2metric}
\begin{split}
ds^{2}(M_6) &=  \frac{e^{-6 \lambda}}{3}(a_1-y^{2})d{s}^{2}(S^{2}_{(1)})
+\frac{e^{-6 \lambda}}{3}(a_2-{y}^{2})d{s}^{2}(S^{2}_{(2)}) \\
&+ e^{- 6 \lambda} \sec^{2} \zeta d {y}^{2} + \frac{1}{9}
\cos^{2} \zeta D\psi^2,
\end{split}
\end{equation}
where $ds^2(S^2_{(1)})$ and $ds^2(S^2_{(2)})$ are both canonical unit radius metrics on
two-spheres. Also, again, $D\psi=d\psi+P$
with $0\le\psi\le 2\pi$ and now $dP=vol(S_{(1)}^2)+vol(S^{2}_{(2)})$.
We also have
\bea
e^{6 \lambda} &=& \frac{2({y}^{2}-a_1)({y}^{2}-a_2)}{2{y}^{2} +cy+ a_1 +
a_2},\nn
\cos^2\zeta&=&\frac{-3 {y}^4-2cy^3-3(a_1+a_2) {y}^{2} + a_1a_2}{(y^2-a_1)(y^2-a_2)} .
\eea

The four-form flux is given by
\begin{eqnarray}
\frac{1}{L^3}G_{(4)} &=& \frac{1}{18 ({y}^{2} -a_1) ({y}^{2}
-a_2)} \left[4 {y}^{5} + 3cy^4+4 {y}^{3} (a_1 + a_2)-cy^2(a_1+a_2) \right. \nonumber  \\
&& \qquad \qquad \left.  - 2 {y}(a_1^2+a_2^2+4 a_1a_2)-ca_1a_2 \right] vol(S^{2}_{(1)})
vol(S^{2}_{(2)}) \nonumber \\
&& + \frac{[{y}^{4} - {y}^{2} ( a_2 + 5 a_1) -2a_1cy- a_1a_2 -
2a_1^2] }{ 9 ({y}^{2} - a_1)^{2}} vol(S^{2}_{(2)}) d {y}
D\psi \nonumber  \\
&& + \frac{[y^{4} - {y}^{2} ( a_1 + 5a_2 )-2a_2cy - a_1a_2 -
2a_2^2] }{ 9({y}^{2} - a_2)^{2}} vol(S^{2}_{(1)}) d {y}
 D\psi .
\end{eqnarray}
Observe that when $a_1=a_2=a$ the solution is the same as the K\"ahler-Einstein
class when $KE_4$ is taken to be $S^2\times S^2$.

For simplicity, in this section we shall again restrict to the $c=0$ class of solutions.
After a rescaling the solution only depends on the ratio
$z=a_1/a_2$. Without loss of generality we can set
\be
a_1=z,\qquad a_2=1
\ee
and take $0<z\le 1$.
The range of $y$ is now easily determined from the zeroes of $\cos^2\zeta$
and we  find
\begin{equation}
\label{just+}
{y}_{1,2} = \mp \left[-\frac{1}{2}(z+1) + \frac{1}{6}\sqrt{X}\right]^{1/2}.
\end{equation}
where $X = 9z^{2} + 30z+9$.

To implement the flux quantisation we will consider the flux through
the cycles $C_1, C_2$, which are the obtained by considering the
$S^2$ fibration over $S^2_{(1)}$ (at a fixed point on $S^2_{(2)}$)
and $S^2_{(2)}$ (at a fixed point on $S^2_{(1)}$), respectively, and
$C_{N}$, which is the four-cycle sitting at the north pole, $y=y_2$,
of the two-sphere fibre. We find
\begin{eqnarray}
\label{hold1}
N_{C_1} &=& -\frac{{y}_{2}L^{3}}{18 \pi l_{11}^3}\left[ 5 +3 z + \sqrt{X}\right], \\
N_{C_2} &=& -\frac{{y}_{2}L^{3}}{18 \pi l_{11}^3
}\frac{\left[ 5z+3 + \sqrt{X}
\right]}{z}, \\
\label{hold2}
N_{C_N} &=& -\frac{{y}_{2}L^{3}}{18 \pi l_{11}^3}\frac{\left[X
  + 3(1+z)
\sqrt{ X}
\right]}{3z} .
\end{eqnarray}
Note that only two of the above charges are independent since
$N_{C_1} + N_{C_2} =  N_{C_N}$.
Taking the ratio of (\ref{hold1}) to (\ref{hold2}) we get the
expression
\begin{equation}
\frac{N_{C_2}}{N_{C_1}} = \frac{1}{z}\frac{\left[ (5 z + 3 )
+ \sqrt{ X} \right]}{\left[ (5  + 3 z) + \sqrt{X}
\right]},
\end{equation}
which can be made a rational number $w$ by choosing $z$
such that
\begin{equation}
z = \frac{2 - w + 2 w^2 - 2
\sqrt{1-w-w^3+w^4}}{3w}.
\end{equation}
The range of $z\in (0,1]$ is
covered once if we impose $w\in[1,\infty)$.
Let us write $w=p/q$ for positive integers $p$ and
$q$ with no common factors and
$p>q$ (we also allow $p=q=1$) and hence
\be
z=\frac{2q^2-qp+2p^2-2(p-q)\sqrt{q^2+qp+p^2}}{3pq} .
\ee
We then choose the length scale $L$ via
\be
\frac{{y}_{2}L^{3}}{18 \pi l_{11}^3}\left[ 5 +3 z + \sqrt{X}\right]=qN .
\ee
This implies that $N_{C_1}=-qN$, $N_{C_2}=-pN$ and $N_{C_N}=-(p+q)N$ are indeed
all integers.

The central charge can now be calculated and we find that it can be written as
\begin{equation}
\label{a1a2c}
c = \frac{3^{3/2}}{2^6}\frac{\left[9(z+1)^3-(3z^2+4z+3){\sqrt X}\right]}
{z^{3/2}}(pq)^{3/2}N^3 .
\end{equation}
As a check on this result, we should be in a position to rederive the
earlier central charge for the $KE_{4}$ base in the special case where
the base is $S^{2} \times S^{2}$. Setting $z=1$ ($p=q=1$), $m=2$ and $M=8$ we
do indeed find agreement.

\subsection{$\mathbf{B_4= S^{2} \times T^{2}}$ and $\mathbf{Y^{p,q}}$:}
We shall now consider the case when the base space $B_4=T^2\times S^2$.
Recall that after dimensional reduction and T-duality, these solutions
gave rise to the $AdS_5\times Y^{p,q}$ solutions of type IIB string
theory where $Y^{p,q}$ are Sasaki-Einstein spaces. The analysis of
\cite{gmsw2} of the regularity of the $Y^{p,q}$ spaces and the
quantisation of the five-form flux can be translated into
the M-theory setting. We now present some of the details.

The metric is given by
\begin{equation}
\label{ypqmetric}
ds^{2}(M_6)= {e^{- 6 \lambda}} d s^{2}(T^{2}) + \frac{1-cy}{6}ds^{2}(S^{2})+{e^{-6 \lambda }} \sec^{2} \zeta dy^{2}
+ \frac{1}{9} \cos^{2} \zeta D \psi^{2},
\end{equation}
where $ds^2(S^2)$ is the metric on a unit radius two-sphere and
$ds^2(T^2)$ is a metric on a two-torus. As above,
$\psi$ has period $2\pi$ and
$D\psi=d\psi+P$ now with $dP=vol(S^2)$ .
We also have
\begin{eqnarray}
e^{6 \lambda} &=& \frac{2 (a - y^{2})}{1 - cy}, \\
\cos^{2} \zeta &=& \frac{a-3 y^{2} + 2 cy^{3}}{a - y^{2}},
\end{eqnarray}
where $a\in(0,1)$ is constant. Here we will only consider $c\ne 0$ and
so without loss of generality we can set $c=1$.
The range of $y$ is given by
$y_{1} \leq y \leq y_{2}$ where $y_1$ and $y_2$ are
the smallest roots of the cubic
$a- 3 y^{2} + 2 y^{3} = 0$.

The four-form flux is given by
\begin{equation}
\label{4formypq}
\begin{split}
\frac{1}{L^3}G_{(4)} &= \frac{-2 y + y^{2}  +  a}{ 6 (a - y^{2})} vol(S^{2}) vol(T^{2}) -
\frac{2(1-y)}{9 } dy  D \psi  vol(S^{2}) \\ & - \frac{a + y^{2} - 2 a y}{3
(a-y^{2})^{2}} dy  D \psi  vol(T^{2}).
\end{split}
\end{equation}

It was shown in \cite{gmsw2} that the parameter $a$ is fixed in terms
of two relatively prime integers $p$ and $q$, $p>q>0$ via:
\be
a=\frac{1}{2}+\frac{3q^2-p^2}{4p^3}{\sqrt{4p^2-3q^2}} .
\ee
The explicit expressions for the roots $y_i$ are
then given by
\bea
y_1=\frac{1}{4p}\left(2p-3q-{\sqrt{4p^2-3q^2}}\right) , \nn
y_2=\frac{1}{4p}\left(2p+3q-{\sqrt{4p^2-3q^2}}\right) .
\eea

The $Y^{p,q}$ spaces are $U(1)$ bundles over a four-dimensional base-space
with Chern-numbers given by $p$ and $q$. The size of the $S^1$ fibre is
 $2\pi l$, which means that the size of the M-theory two-torus
is $4\pi^2(l_{11}/L)^3/l$ (see, e.g., section 6 of \cite{gmmw3}).
Furthermore, the overall length scale of the type IIB solution translates
into the parameter $L$ of the $D=11$ solution taking
the form \be \frac{L^3}{12\pi
l_{11}^3}=\frac{p^2}{q\left[2p+{\sqrt{4p^2-3q^2}}\right]}N , \ee where
$N$ is an arbitrary integer. Using this we can write
\be vol(T^2)=\pi
\frac{q^2\left[p+\sqrt{4p^2-3q^2}\right]}{p^2N}.
\ee

We now confirm that these constraints ensure that the four-form flux
is properly quantised. If we let $C_{S^2}$ and $C_{T^2}$ denote the
four-cycles given by the $S^2$ fibration over the $S^2$ and $T^2$,
respectively, and $C_N$ the four-cycle given by the north pole of
the fibration, $y=y_2$, we find \bea N_{C_{S^2}}&=&-N , \nn
N_{C_{T^2}}&=&-p , \nn N_{C_N}&=& q-p . \eea
In addition we have
$N_{C_S}=(q+p)$. To compare with the IIB results, one should choose
a different basis of four-cycles. In particular, instead of
$C_{T^2}$ and $C_N$ we choose $(1/2)(C_S-C_N)$ and $(1/2)(C_S+C_N)$
(these are the four-cycles $C_1\times T^2$ and $C_2\times T^2$ where
the $C_i$ are the basis of two-cycles on the $S^2$ fibration over
$S^2$ that were discussed in \cite{gmsw2}). We find that the
corresponding fluxes are given by $p$ and $q$, respectively.

Calculating the central charge using \reef{ccharge2} we get
\begin{eqnarray}
\label{6dimc}
c &=&  \frac{3 p^{2}[3q^{2}-2p^{2} + p\sqrt{4 p^{2} -3
q^{2}}]}{4q^{2} [ 2 p + \sqrt{4p^{2}-3q^{2}} ]}N^2  \nonumber \\
&=& \frac{\pi^{3} N^{2}}{4 Vol(Y^{p,q})} ,
\end{eqnarray}
in agreement with \cite{gmsw2} as expected.
Note that in contrast to the $B_4$ considered in the last two subsections,
when $B_4=S^2\times T^2$, not all of the fluxes scale with $N$; this
leads to the $N^2$ scaling behaviour for the central charge with fixed $p$ and $q$,
rather than the $N^3$ behaviour we saw previously.

\section{Supersymmetric Membranes and Chiral Primaries}

We now consider probe membranes wrapping supersymmetric cycles on $M_6$.
These configurations correspond to chiral primaries
in the dual conformal field theories. For the most general class of solutions
of \cite{gmsw1} we claim that the conformal dimension of these operators can be obtained from the formula
\begin{equation}
\label{M2R}
\Delta(\Sigma_2) =  \tau_{M2} L^3\int_{\Sigma_2} e^{3 \lambda}
vol_{M_{6}}(\Sigma_2) .
\end{equation}
Here $vol_{M_6}(\Sigma_2)$ is the volume form of the cycle induced from the metric $ds^2(M_6)$ and
$\tau_{M2}$ is the tension of the membrane and is given in our conventions by
$\tau_{M2}=1/(4\pi^2l_{11}^3)$. That the volume of the wrapped brane (suitably dressed
by the warp factor) should be associated with the conformal dimension of the operators
rather than the mass of the operators was explained in the context of Sasaki-Einstein compactifications
in type IIB supergravity in \cite{baryonspec}. In the next section we will confirm that
this is indeed the correct formula, by showing that the conformal dimension is related to the $R$-charge
of these operators by the correct formula, $\Delta=(3/2)R$.

In order to be supersymmetric, the membranes must wrap two-cycles in $M_6$ that are
calibrated by a certain two-form (generalised) calibration \cite{gpt}.
Recall \cite{gmsw1} that the $D=11$ Killing spinors of the most general solutions \reef{warpedansatz}
have the form
\begin{equation}\label{ksp}
\epsilon = \psi\otimes e^{\lambda / 2}  \xi ,
\end{equation}
where $\psi$ is a Killing spinor on $AdS_5$ and $\xi$ is a non-chiral spinor on $M_6$.
The $SU(2)$ structure specified by $J$, $\Omega$, $K^1$, $K^2$ and $\cos\zeta$ can be
constructed from bi-linears of $\xi$ as explained in \cite{gmsw1}. Continuing to use the notation
of \cite{gmsw1}, let us focus on the bi-linear two-form given by
\bea\label{formypr}
Y'&=&\frac{1}{2}\bar\xi\gamma_{(2)}\gamma_7\xi\nn
&=&-\sin\zeta J + K^1\wedge K^2 .
\eea
Using the results of \cite{gmsw1} a short calculation now shows that
$Y'$ always satisfies the generalised calibration condition
\be\label{gencally}
d(e^{3\lambda}Y')=-i_{\tilde K^2}\frac{G_{(4)}}{L^3}
\ee
where $\tilde K^2=\cos\zeta K^2$ is the one-form whose dual vector on $M_6$
is Killing (in the coordinates in \reef{anexc} it is $\tilde K^2=3\partial_\psi$).
We now claim that if a membrane wraps a two-cycle $\Sigma_2$ on $M_6$ which is calibrated by
$Y'$, i.e.
\be\label{calcondm}
vol_{M_6}(\Sigma_2)=Y'|_{\Sigma_2} ,
\ee
then it is a supersymmetric configuration.

One way to see why $Y'$ is the relevant object is to consider the
two-form $\bar\Omega$, the membrane generalised calibration, of the
full $D=11$ solution. Recall that for any supersymmetric solution of $D=11$ supergravity with
a Killing spinor $\epsilon$ one can construct the bilinears $\bar K$, a one-form, and $\bar\Omega$, a two-form
(there is also a five-form that is not relevant for the present discussion). It was shown in
\cite{gp} that
$\bar\Omega$ satisfies the generalised calibration condition $d\bar\Omega=i_{\bar K} G_{(4)}$.
Next, restricting to the class of supersymmetric $AdS_5$ solutions of interest,
we can use the decomposition of the $D=11$ Killing spinor \reef{ksp} to show that restricting to directions
tangent to $M_6$, we have $\bar\Omega\to 2e^{\lambda}\bar\psi\psi Y'$ and $\bar K\to -2e^{\lambda}\bar\psi\psi\tilde K^2$.
Properly taking into account the conformal factors then leads to \reef{gencally}.

We shall now give detailed expressions for the calibration two-form $Y'$ \reef{formypr} for
the explicit solutions of \cite{gmsw1} that we considered in the last section,
and then use this to elucidate the corresponding calibrated cycles satisfying \reef{calcondm}.
As an extra check that we have indeed properly identified supersymmetric cycles, we have
carried out a direct analysis of the $\kappa$-symmetry conditions in appendix A (similar calculations
have been carried out in the context of Sasaki-Einstein solutions in \cite{Arean:2004mm,Canoura:2005uz}).
Having identified the supersymmetric cycles, the dimension of the corresponding chiral operators
is determined using \reef{M2R}.

\subsection{$\mathbf{B_4= KE_4}$:}
For this case the calibration two-form \reef{formypr} is given by
\begin{equation}
\label{Y'KE4}
Y' = -\frac{e^{-6 \lambda}}{3} (b-y^{2})\sin\zeta J_{KE} +
\frac{e^{-3\lambda}}{3} d y D\psi .
\end{equation}

We now consider a membrane wrapping the two sphere fibre at a fixed
point on the $KE_4$. From \reef{KEmetric} the volume form for this
cycle is simply $(e^{-3\lambda}/3)dy d\psi$ and hence it is
calibrated by $Y'$. The dimension of the dual chiral primaries is
given by \reef{M2R} and, for the class of solutions with $c=0, b=1$
for which we carried out the flux quantisation in the last section,
we find \be \Delta=6(2-\sqrt{3})\frac{N}{h} . \ee

We can also consider a membrane wrapping a two-cycle on $KE_4$ sitting at one of the
poles of the $S^2$ fibration, $y=y_1, y_2$. Since at the poles we have $\sin\zeta=\pm 1$ it
is clear that a holomorphic two-cycle $C_i$ on $KE_4$, i.e. one that is calibrated by $J_{KE}$,
at $y=y_1,y_2$ is calibrated by $Y'$. We can expand such a cycle in the basis $\Sigma_a$ of $H_2(KE_4,\bbZ)$
that we introduced in section 2.1 as $C_i=C_i^a\Sigma_a$ where $C_i^a\in \bbZ$.
For membranes wrapping these cycles, in the case that $c=0,b=1$, we then find that
\be
\Delta=3(\sqrt{3}-1)\frac{C_i^amn_a}{h}N.
\ee
The fact that the $\Delta$ is the same for either pole is because
$y_2=-y_1$.

\subsection{$\mathbf{B_4= S^{2} \times S^{2}}$:}
For this case we have
\begin{equation}
\label{Y'product}
 Y' =
-\frac{e^{-6\lambda}}{3}(a_1-y^2)\sin\zeta vol(S^2_{(1)})
-\frac{e^{-6\lambda}}{3}(a_2-y^2)\sin\zeta vol(S^2_{(2)})
+\frac{e^{-3\lambda}}{3 } d y  D\psi.
\end{equation}

The two-sphere fibre is again calibrated by $Y'$. When $c=0, a_1=z, a_2=1$, membranes
wrapping this cycle gives rise to chiral primaries with
\be
\Delta=\frac{3q}{8}(5+3z-\sqrt{X})N .
\ee
In addition the two-spheres on the base, $S^2_{(1)}$ or $S^2_{(2)}$, located at either
$y=y_1$ or $y=y_2$ are also calibrated by $Y'$. The
conformal dimensions corresponding to the two-cycles on the base at either pole are the same and
are given, when $c=0, a_1=z, a_2=1$, by
\bea
\Delta(S^2_{(1)})&=&\frac{3q}{8}(-3z-1+\sqrt{X})N , \nn
\Delta(S^2_{(2)})&=&\frac{3q}{16z}(-3z^2-8z+3+(1+z)\sqrt{X})N ,
\eea

\subsection{$\mathbf{B_4= S^{2} \times T^{2}}$ and $\mathbf{Y^{p,q}}$:}
This case provides a good check on our formulae since we can compare with
the dimension of the chiral primaries that have already been calculated
in the context of type IIB supergravity using the Sasaki-Einstein spaces $Y^{p,q}$ .
For this case we find that the calibration two-form is given by
\begin{equation}
\label{Y'productT2}
 Y' =
-\frac{e^{-6\lambda}}{3}(a-y^2)\sin\zeta vol(S^2)
-e^{-6\lambda}\sin\zeta vol(T^2)
+\frac{e^{-3\lambda}}{3 } d y  D\psi.
\end{equation}

The two-sphere fibre is once again calibrated by $Y'$ and the corresponding conformal
dimension is given by
\be
\Delta=\frac{p}{q^2}(2p-\sqrt{4p^2-3q^2})N .
\ee
After dimensional reduction and $T$-duality the wrapped membrane corresponds to a D3-brane
wrapping the three-cycle given by fixing a point on the round two-sphere in the $Y^{p,q}$ space.
Our result for $\Delta$ agrees with that given in \cite{Newcycles}.

We next consider membranes wrapping the two-sphere by fixing a point on the $T^2$
and fixing $y=y_1$ or $y=y_2$. These are calibrated by $Y'$ and we find
\bea
\Delta(S^2(y_1))&=&\frac{1}{2q^2}\left( -4p^2+2pq+3q^2+(2p-q)\sqrt{4p^2-3q^2}\right)N , \nn
\Delta(S^2(y_2))&=&\frac{1}{2q^2}\left( -4p^2-2pq+3q^2+(2p+q)\sqrt{4p^2-3q^2}\right)N .
\eea
After dimensional reduction and $T$-duality these wrapped membranes correspond to D3-branes
wrapping three-cycles in $Y^{p,q}$ that were discussed in \cite{toricgeom}
and our formula for $\Delta$ is in exact agreement.

We can also consider membranes wrapping the $T^2$ by fixing a point
on the $S^2$ and fixing $y=y_1$ or $y=y_2$. These are also
calibrated by $Y'$ and we find \bea
\Delta(T^2(y_1))&=&\frac{p}{2q}(3q+2p-\sqrt{4p^2-3q^2}) , \nn
\Delta(T^2(y_2))&=&\frac{p}{2q}(3q-2p+\sqrt{4p^2-3q^2}) . \eea After
dimensional reduction and $T$-duality these wrapped membranes
correspond to momentum waves around the $\alpha$ direction in the
Sasaki-Einstein picture. In fact these are precisely the conformal
dimensions of the long BPS mesonic operators ${\cal L}_{\pm}$ of the dual CFT that
were noted in \cite{Benvenuti:2005cz} where they were also identified with massless
geodesics in the $AdS_5\times Y^{p,q}$ solutions.

\section{Consistent Truncation}

In this section we show that for the most general class of supersymmetric $AdS_5$ solutions of \cite{gmsw1}
there is a consistent truncation to minimal five-dimensional gauged supergravity.
We shall argue that the abelian gauge field of the gauged supergravity theory can be
identified with the $R$-symmetry of the dual SCFT. We use this to determine the $R$-charges
of the chiral primaries dual to the wrapped branes discussed in the last section and demanding that
$\Delta=3/2 R$ we derive the formula for $\Delta$ presented in \reef{M2R}. Following this we discuss some aspects of the fluid-dynamics of
the thermal plasma of the SCFTs with non-zero chemical potential.

\subsection{The ansatz}
The ansatz for the $D=11$ metric is given by
\reef{warpedansatz} where we replace the $AdS_5$ metric with an arbitrary
$D=5$ metric: 
\be \label{ctmetans}
ds^2=L^2e^{2\lambda}[ds^2_5+ds^2(M_6)]
\ee
with
\bea\label{go1}
ds^2(M_6)&=&g^4_{ij}(x,y)dx^idx^j+e^{-6\lambda}\sec^2\zeta dy^2+\frac{1}{9}\cos^2\zeta(d\psi+\rho+A)\nn
&=&e^ie^i +(e^5)^2+(e^6)^2 ,
\eea
where $e^5 \equiv K^1=e^{-3\lambda}\sec\zeta$ and $e^6=K^2+(1/3)\cos\zeta A$ and $i,j=1,\dots,4$.
Here we have just made the shift $d\psi\to d\psi+A$ in the general metric \reef{anexc} of \cite{gmsw1}.

The ansatz for the four-form is much less obvious. After some trial and error we find
that it is given by
\be\label{gogo}
\frac{1}{L^3}G_{(4)}=d\bar C +\frac{1}{3}dy [ (*_5F)+\frac{1}{3}A  F]
\ee
where
\be\label{ceeyp2}
\bar C=C^0+\frac{1}{3}A  (e^{3\lambda}Y')
\ee
and $L^3dC^0$ is the four-form flux of the undeformed solution of \cite{gmsw1} given below.
After a little reorganisation,
one finds that
\be \label{gogoreorg}
\frac{1}{L^3}G_{(4)}=g+\frac{e^{3\lambda}}{3}(-\sin\zeta J+K^1 e^6)F+\frac{1}{3}e^{3\lambda}\cos\zeta K^1 (*_5F)
\ee
where $g$ is the four-form flux given in \cite{gmsw1} after the substitution
$d\psi\to d\psi+A$:
\bea
g&=&-\frac{1}{2}e^{12\lambda}\partial_y(e^{-6\lambda})J^2
-e^{-3\lambda}\sec\zeta(*_4d_4e^{6\lambda})K^1\nn
&&-\frac{1}{3}e^{6\lambda}\cos^3\zeta(*_4\partial_y\rho)e^6+e^{3\lambda}[\frac{1}{3}\cos^2\zeta(*_4 d_4\rho)-4J]K^1 e^6
\eea

We now substitute this ansatz into the $D=11$ equations of motion:
\bea
R_{\mu\nu}-\frac{1}{12}(G_{\mu\sigma_1\sigma_2\sigma_3}G_{\nu}{}^{\sigma_1\sigma_2\sigma_3}-\frac{1}{12}g_{\mu\nu}G^2)=0 , \nn
dG_{(4)}=0,\qquad d*G_{(4)}+\frac{1}{2}G_{(4)}\wedge G_{(4)}=0 , \eea
where
$G^2=G_{\sigma_1\sigma_2\sigma_3\sigma_4}G^{\sigma_1\sigma_2\sigma_3\sigma_4}$.
After some calculation\footnote{The Bianchi identity for $G_4$ is
simple to verify. In checking the equations of motion for $G_4$ we
used equations (2.47) and (2.49) of \cite{gmsw1} and we also used
the exterior derivative of $e^{6\lambda}$ times equation (2.16) of
\cite{gmsw1}. In checking the Einstein equations we used the fact
that the $AdS_5$ is a solution and focused on the $F$ dependent
pieces.} using the conditions in \cite{gmsw1}, we deduce that the
$D=5$ metric $g$ and the two-form field strength $F=dA$ must satisfy
\bea\label{eomgs}
R_{ab}=-4g_{ab}+\frac{1}{6}\left[F_{ac}F_{b}{}^{c}-\frac{1}{6}g_{ab}F^2\right] , \nn
d(*_5F)+\frac{1}{3}F\wedge F=0 , \eea where $F^2=F_{ab}F^{ab}$. These
equations can be derived from an action whose lagrangian is given by
\be\label{lagmgs}
\sqrt{-g}\left[R+12-\frac{1}{12}F^2+\frac{1}{108}\epsilon^{a_1a_2a_3a_4a_5}F_{a_1a_2}F_{a_3a_4}A_{a_5}\right]
\ee which is precisely the bosonic Lagrangian of minimal $D=5$
gauged supergravity.

We have thus shown that our ansatz is a consistent Kaluza-Klein truncation at the level of bosonic fields.
In particular, any bosonic solution of minimal $D=5$ gauged supergravity can be uplifted using
the ansatz to give a solution of $D=11$ supergravity that is based on an arbitrary
general supersymmetric $AdS_5$ solution of \cite{gmsw1}. In appendix B we show that
the Kaluza-Klein truncation is also consistent after including the fermions.

\subsection{$R$-charge}
Recall that the Killing vector $\partial_\psi$ arises in all supersymmetric
$AdS_5$ solutions of $D=11$ supergravity and hence it must be related to
the $R$-symmetry in the dual SCFT. Clearly the ansatz for consistent truncation above
involves gauging shifts of the coordinate $\psi$ and so it is natural to
identify the gauge field $A$ with the $R$-symmetry.

While we claim that this is indeed the right interpretation it is
worth mentioning a subtlety. In general the $R$-symmetry should be
related to a linear combination of the isometry generated by the
Killing vector $\partial_\psi$ with gauge transformations of the
$D=11$ three-form potential $C_{(3)}$ which satisfies
$dC_{(3)}=G_{(4)}$. This can be seen in detail in the context of the
$B_4=S^2\times T^2$ solutions. After dimensional reduction and
$T$-duality the Killing vector $\partial_\psi$ does not become the
Reeb Killing vector of the Sasaki-Einstein space itself (which is
known to be dual to the $R$-symmetry) but rather a linear
combination of the Reeb vector with another Killing vector. The
gauge transformations of $C_{(3)}$ account for the latter Killing
vector. In fact, we can be more explicit and show that if we
dimensionally reduce our ansatz above then we recover the ansatz for
the consistent truncation of the $AdS_5\times Y^{p,q}$ solutions to
metric plus $R$-symmetry gauge field that was discussed in
\cite{Buchel:2006gb}. This is carried out in appendix C.

We now want to calculate the $R$-charges of the chiral primaries dual
to the wrapped membranes that we considered in section 3. In the Lagrangian for a
membrane wrapping the supersymmetric cycle $\Sigma_2$, the Wess-Zumino term gives rise to the term
\begin{equation}
\label{Rvol}
\tau_{M2} \int_{\Sigma_{2}} C_{(3)} = \frac{1}{2}RA.
\end{equation}
In analogy with the way in which a classical Lagrangian particle of charge $q$ travelling in an
electromagnetic field couples through a four-velocity term $q A_{\mu} v^{\mu}$, we deduce that
the wrapped branes carry $R$ units of $R$-charge. The factor of 1/2 appearing in \reef{Rvol}
comes from properly normalising the gauge field as we discuss in appendix C. In particular
$A/2$ is the natural normalisation for the gauge field.

Now consider the expression for $C_{(3)}$ in the consistent truncation ansatz \reef{gogo}, \reef{ceeyp2}.
Focussing on the terms that are linear in $A$ we get
\bea
\tau_{M2} \int_{\Sigma_{2}} C_{(3)}
&=& \frac{1}{3}A \tau_{M2}L^3\int_{\Sigma_2} e^{3\lambda}Y'\nn
&=&\frac{1}{3}A
\tau_{M2}L^3\int_{\Sigma_2} e^{3\lambda}vol_{M_6} \ ,
\eea
where we used the calibration condition
\reef{calcondm}. We thus deduce that the $R$-charge
of the supersymmetric wrapped membranes are given by
\be
R=\frac{2}{3}\tau_{M2}L^3\int_{\Sigma_2} e^{3\lambda}vol_{M_6} \ .
\ee
Since for chiral primaries we have
$\Delta=\frac{3}{2}R$, this provides a derivation of our formula
for $\Delta$ given in \reef{M2R}.

\subsection{Thermal Plasma}
The consistency of our Kaluza-Klein truncation implies, by definition,
that any solution of minimal gauged supergravity
can be uplifted to eleven dimensions. In particular, black hole solutions
of minimal gauged supergravity with non-vanishing $R$-charge can be uplifted
and this allows us to extract non-trivial information about
the hydrodynamics of the strongly coupled thermal plasma of the dual field theory.

In fact the relevant calculations were all carried out in detail in
\cite{Buchel:2006gb} (these calculations were extended from minimal gauged supergravity to
a more general class of theories in \cite{Benincasa:2006fu}). One first extracts the thermodynamical
quantities of the black hole solutions using the technique of
\cite{Balasubramanian:1999re}. To study the hydrodynamics of the
thermal plasma dual to the black holes, one needs to calculate the
retarded Green's function of the boundary stress tensor using the prescription
of \cite{Son:2002sd}. The shear viscosity $\eta$ can then be obtained from a Kubo relation.

Using this analysis we conclude that for the most general supersymmetric $AdS_5$ solutions of
M-theory, the thermal plasmas of the dual CFTs with
non-zero $R$-charge chemical potential
all have the property that
\be
\frac{\eta}{s}=\frac{1}{4\pi} .
\ee
In addition, again using the results of \cite{Buchel:2006gb}, we can conclude that
the speed of sound in the thermal plasma is $1/{\sqrt 3}$.

\section{Discussion}
In this paper we have obtained some results about the CFTs dual to
supersymmetric $AdS_5$ solutions that were analysed in \cite{gmsw1}.
In section 2, for some explicit solutions,
we explicitly calculated the central charges of the dual CFTs. For
the solutions with $B_4=KE_4$ or $B_4=S^2\times S^2$ the calculations were performed
in the special case that $c=0$. It would be interesting to extend this to $c\ne 0$
but repeating the same steps is messy since the range of
the coordinate $y$ is then given by the roots of a quartic, rather than,
effectively, a quadratic.

In section 3, for the most general class of $AdS_5$ solutions of \cite{gmsw1}
we found the conditions for wrapped membranes
to be supersymmetric by elucidating the appropriate
generalised calibration two-form which can be constructed from the Killing spinors.
For the explicit solutions we showed that various two-cycles are supersymmetric and then
calculated the conformal dimensions of the corresponding chiral
primaries.

In section 4 we showed that for the most general solutions of \cite{gmsw1}
there is a consistent Kaluza-Klein truncation to minimal gauged supergravity
in five dimensions. The gauge field corresponds to the abelian $R$-symmetry
of the $N=1$ SCFT.
We used the consistent truncation to determine some properties of the hydrodynamics
of the dual CFT using the results of \cite{Buchel:2006gb}. In particular, we extended
the domain of validity of the universal result that $\eta/s=1/(4\pi)$.

The fact that the Kaluza-Klein reduction is a consistent
truncation is somewhat surprising.
For cases in which the solutions have additional
isometries, we do not expect, in general, to be able to extend the
result to include more gauge fields (e.g. see \cite{pope}).
However, for the special class of solutions with
$N=2$ supersymmetry contained in \cite{gmsw1} (and further analysed in \cite{llm})
it seems plausible that there is a consistent truncation that maintains
the whole $SU(2)\times U(1)$ $R$-symmetry.
More generally it is natural to conjecture that the most general supersymmetric $AdS_n$ solutions
always admit consistent truncations to a gauged supergravity that contain at
least some if not all of the $R$-symmetry. It would be interesting to
investigate this conjecture using the classification
results of \cite{Gauntlett:2006ux}.

\section*{Acknowledgements}
We would like to thank Alex Buchel and Dan Waldram for helpful
discussions. JPG is supported by an EPSRC Senior Fellowship and a
Royal Society Wolfson Award. JPG would also like to thank the
Aspen Center for Physics where some of this work was done. OV
wishes to thank the Generalitat Valenciana for the FPI research
fellowship that he enjoyed whilst some of this work was carried
out, and Imperial College for its warm hospitality during his visit
at the early stages of this work.

\appendix

\section{$\kappa$-symmetry}

The condition for a probe membrane to be supersymmetric is that it
satisfies the condition \cite{Becker:1995kb}
\begin{equation}
\Gamma_{\kappa} \epsilon = \epsilon,
\end{equation}
Here $\epsilon$ is a $D=11$ Killing spinor and the $\kappa$-symmetry
projection matrix $\Gamma_\kappa$ is given by
\begin{equation}
\Gamma_{\kappa} = \frac{1}{3! \sqrt{-g}} \epsilon^{u_{1} u_{2}
u_{3}}{\sigma_{u_{1}u_{2} u_{3}}},
\end{equation}
where here $g$ is the determinant of the induced world-volume metric and
$\sigma_{u}$ are induced world-volume gamma matrices: if the
membrane is defined by the maps $x^\mu(\xi)$, we have
$\sigma_u=\partial_u x^\mu \Gamma_\mu$ where $\Gamma^\mu$ are the
$D=11$ Gamma matrices.

For the general $D=11$ solutions of \cite{gmsw1} the $D=11$ Killing spinors
take the form
\begin{equation}
\label{11dkill}
\epsilon = \psi \otimes e^{\lambda / 2}  \xi,
\end{equation}
where $\psi$ is a Killing spinor on $AdS_5$ and $\xi$ is
a spinor on $M_6$ satisfying differential equations written in \cite{gmsw1}.
The $D=11$ gamma matrices can be decomposed as
\begin{eqnarray}
\Gamma^{{a}} &=& \rho^{a} \otimes \gamma_{7} , \nonumber \\
\label{cliff}
\Gamma^{{m}} &=& 1 \otimes \gamma^{m},
\end{eqnarray}
where $a,b = 0,1,...,4$ and $m,n = 1,2,...,6$ are frame indices on
$AdS_{5}$ and $M_{6}$ respectively, and we have
\begin{equation}
\begin{array}{cc}
[\rho^{a}, \rho^{b} ]_{+} = -2 \eta^{ab}, &
[\gamma^{m}, \gamma^{n} ]_{+} = 2 \delta^{mn},
\end{array}
\end{equation}
with $\eta^{ab} = $ diag($-1,1,1,1,1$) and $\rho_{01234} =-1$.

It will be useful to have an explicit expression for the Killing spinors $\psi$.
In global co-ordinates the metric on $AdS_5$ can be written as
\begin{equation}
ds^{2} (AdS_{5}) = [ -\cosh^{2} r d t^{2} + d
r^{2} + \sinh^{2} r d \Omega^{2}_{3} ],
\end{equation}
where
\begin{equation}
d \Omega^{2}_{3} = (d \alpha_{1})^{2} + \sin^{2} \alpha_{1} \left[ (d
\alpha_{2})^{2} + \sin^{2} \alpha_{2} (d \alpha_{3})^{2} \right],
\end{equation}
is the metric of a unit three-sphere parametrised by the three angles ($\alpha_{1}, \alpha_{2} ,
\alpha_{3}$),
with $0 \leq \alpha_{1}, \alpha_{2} \leq \pi$ and $ 0 \leq \alpha_{3}
\leq 2 \pi$.
The Killing spinors satisfy
\be
D_a\psi\equiv(\partial_a-\frac{1}{4}\omega_{abc}\rho^{bc})\psi=\frac{i}{2}\rho_a\psi
\ee
and, in the obvious orthonormal frame, take the explicit form (see \cite{Lu:1998nu,Grisaru:2000zn})
\begin{equation}
\label{adsspin}
\psi = e^{i \frac{r}{2} \rho_{1}}
 e^{i \frac{t}{2} \rho_{0} }
 e^{\frac{\alpha_{1}}{2} \rho_{21}}
e^{\frac{\alpha_{2}}{2} \rho_{32} }
e^{\frac{\alpha_{3}}{2} \rho_{43} }\psi_0,
\end{equation}
where $\psi_0$ is a constant spinor on $AdS_{5}$.

It will also be useful to recall \cite{gmsw1} that
the spinor $\xi$ on $M_6$ can be written
\begin{equation}
\xi = \sqrt{2} \cos \alpha \eta_{1} +
\sqrt{2} \sin \alpha \eta^{*}_{2},
\end{equation}
where $\zeta=\pi/2-2\alpha$, and $\eta_{1}$  and $\eta^{*}_{2}$ are
unit norm chiral spinors on the $M_{6}$ satisfying the projections
\begin{eqnarray}
\label{projections1}
\gamma_{12} \eta_{1} &=& \gamma_{34} \eta_{1} = -\gamma_{56} \eta_{1}
= i \eta_{1} , \\
\label{projections2}
\gamma_{12} \eta^{*}_{2} &=& \gamma_{34} \eta^{*}_{2} = \gamma_{56}
\eta^{*}_{2}
= i \eta^{*}_{2} .
\end{eqnarray}
Here the components $1,2,3,4$ refer to the base of the $SU(2)$
structure defined by $\xi$ and $5,6$ refer to the other two
directions: see \reef{anexc1}, \reef{anexc} (and we refer to
\cite{gmsw1} for more details).

We are now in a position to show that the membranes wrapping
the two-cycles calibrated by $Y'$ that were considered in the
text are in indeed supersymmetric. For simplicity
we only present the details for the solutions with $B_4=KE_4$.

\noindent \textbf{Case 1 : Fibre membrane probe.}
We first consider a probe membrane wrapping the $S^{2}$ fibre parametrised by
$y,\psi$. We therefore simply set $\xi^{u} = (t,y,\psi)$.
Using this embedding and decomposing the $D=11$ gamma-matrices we find that
$\Gamma_{\kappa}$ reduces to:
\begin{equation}
\Gamma_{\kappa} = \rho_0 \otimes \gamma_7\gamma_{56} .
\end{equation}
Since $\gamma_7\gamma_{56}\xi=\xi$ we deduce that the
condition $ \Gamma_{\kappa} \epsilon = \epsilon$
is equivalent to
\begin{equation}
\rho_{0} \psi = \psi.
\end{equation}
If we now return to the explicit expression for the $AdS_{5}$ spinor
(\ref{adsspin}), we see that $\rho_0$ commutes with all matrices on
the right hand side except for $\rho_{1} $. Thus we conclude that
in order to preserve supersymmetry
we must place our membrane probe at $r = 0$,
\textit{i.e.} at the centre of $AdS_{5} $ space.
Note that this condition arises because we have demanded that the wrapped
membrane is independent of $t$.

\noindent \textbf{Case 2 : Base membrane probe.} We now consider
membranes wrapping two-cycles on the $KE_4$ base. More precisely we
show that a membrane wrapping a holomorphic cycle on $KE_4$ that is
located at one of the poles of the fibration $y=y_1,y_2$ and at the
centre of $AdS_5$ is supersymmetric. To see this we set
$\xi^u=(t,\xi^s)$, $s=1,2$, and consider configurations
$x^i=x^i(\xi^s)$, where $x^i$, $i=1,2,3,4$ are coordinates on
$KE_4$. We now find \be \Gamma_{\kappa} = \rho_0 \otimes
\frac{1}{\sqrt{g'}}(\partial_sx^i\partial_s x^j)\gamma_7\gamma_{ij} ,
\end{equation}
where $g'$ is the determinant of the induced metric on the two-cycle
in $KE_4$. Suppose that we are sitting at the pole
$y=y_1$ (the $y=y_2$ case is similar). We then
have $\sin\zeta=-1$ corresponding to $\alpha=\pi/2$ and hence
$\gamma_7\gamma_{12}\xi=\gamma_7\gamma_{34}\xi=\xi$.
If we again demand
\be
\rho_0\psi=\psi ,
\ee
which is achieved by locating the cycle at the centre of $AdS_5$, $r=0$,
we see that the condition $\Gamma_\kappa\epsilon=\epsilon$ is precisely
that for a supersymmetric two-cycle in $KE_4$.
In particular, we must restrict to
holomorphic curves in $KE_4$ in order to preserve supersymmetry.

\section{Fermions and Consistent Truncation}
We now show that we can extend the consistent Kaluza-Klein
truncation to include the fermions. We start with the
variation of the $D=11$ gravitino $\Psi_\mu$:
\be\label{d=11var} \delta \Psi_\mu = \nabla_\mu \epsilon
+\frac{1}{288}(\Gamma_\mu{}^{\nu_1\nu_2\nu_3\nu_4}-8\delta_\mu^{\nu_1}\Gamma^{\nu_1\nu_2\nu_3})
G_{\nu_1\nu_2\nu_3\nu_4} \epsilon  \; ,\ee
and decompose the $D=11$ Killing spinor as $\epsilon = \varepsilon \otimes
e^{\lambda / 2}  \xi$, where $\varepsilon$ is an arbitrary spinor on the external
 $D=5$ space-time and
$\xi$ is the $M_6$ Killing spinor for the undeformed solutions. We will only need below the Kaluza-Klein
ansatz for the  external component of the gravitino; in tangent space, it reads
 $\Psi_a  = \psi_a \otimes e^{-\lambda/2} \xi$, where
$\psi_a$ is the $D=5$ gravitino.

Combining this with the ansatz \reef{ctmetans} for the metric and \reef{gogoreorg} for the
four-form, and substituting
into \reef{d=11var} we find that the components tangent to $M_6$,
{\setlength\arraycolsep{1pt}
\begin{eqnarray} \label{varPsiint}
\delta \Psi_i &=& \delta^0 \Psi_i + \frac{1}{144} F_{ab} \rho^{ab} \varepsilon \otimes
\left( 2 \gamma_7 \gamma_{i5} \cos \zeta
+ \left(\gamma_i{}^{jk} -4\delta_i^j \gamma^k \right) J_{jk} \sin \zeta -2\gamma_{i56} \right)
e^{-\lambda / 2}  \xi , \nonumber \\
\delta \Psi_5 &=& \delta^0 \Psi_5 + \frac{1}{144} F_{ab} \rho^{ab} \varepsilon \otimes
\left( -4 \gamma_7 \cos \zeta
+ \gamma_5{}^{jk} J_{jk} \sin \zeta +4 \gamma_6 \right)
e^{-\lambda / 2}  \xi , \nonumber \\
\delta \Psi_6 &=& \delta^0 \Psi_6 + \frac{1}{144} F_{ab} \rho^{ab} \varepsilon \otimes
\left( 2\left(3-\gamma_7 \gamma_{56} \right) \cos \zeta
+ \gamma_6{}^{jk} J_{jk} \sin \zeta -4 \gamma_5 \right)
e^{-\lambda / 2}  \xi ,
\end{eqnarray}
}vanish identically. Indeed, in \reef{varPsiint}, $\delta^0$ corresponds to the variation of
the undeformed gravitino, which vanishes due to the fact that $\xi$ is the Killing spinor
of the undeformed solutions, and the $F$-dependent terms also vanish
on account of the projections satisfied by $\xi$
that can be read off from \reef{projections1}, \reef{projections2}.

For the external directions $\delta \Psi_a$ of \reef{d=11var} we find, once the projections on $\xi$
have been taken into account,
\begin{eqnarray}
&&  \delta \psi_a \otimes e^{-\lambda/2} \xi  = \left( \nabla_a -A_a \partial_\psi -\frac{1}{24} F_{bc}
\left( \rho_a{}^{bc} + 4 \delta_a^b \rho^c\right) -\frac{i}{2} \rho_a \right) \varepsilon \otimes e^{-\lambda / 2}  \xi
\nonumber \\
&& \qquad  +\frac{1}{2} \rho_a \varepsilon \otimes \left(i + \partial_m \lambda \gamma_7 \gamma^m +
\frac{1}{144} e^{-3\lambda} \gamma_7 \gamma^{m_1 m_2 m_3 m_4} g_{m_1 m_2 m_3 m_4} \right)
e^{-\lambda / 2}  \xi . \qquad
\end{eqnarray}
Using the fact that  $\xi  = e^{\frac{i}{2}\psi} \xi_0$, where $\xi_0$ is independent of
the coordinate\footnote{This follows from equations (2.45), (2.32),
(C.13), (B.6), (B.1) of \cite{gmsw1} and the fact that $\lambda$ is $\psi$-independent.
Here, the $D=5$ gravitino $\psi_a$ should not be confused with the $M_6$ coordinate $\psi$.}
$\psi$, and  equation (2.8) of \cite{gmsw1}, this becomes
\begin{equation}
\delta \psi_a = \left( \nabla_a -\frac{i}{2} A_a  -\frac{1}{24}
\left( \rho_a{}^{bc} + 4 \delta_a^b \rho^c\right)F_{bc} -\frac{i}{2} \rho_a \right) \varepsilon .
\end{equation}
This is precisely the supersymmetry variation of the gravitino in
minimal $D=5$ gauged supergravity corresponding to the Lagrangian whose
bosonic terms are as in \reef{lagmgs}.

\section{Consistent Truncation and Normalisation of $R$-Charge}

For the explicit solution with $B_4=S^2\times T^2$, the Kaluza-Klein ansatz
\reef{go1}, \reef{gogo}, \reef{ceeyp2} takes the detailed form
\begin{equation}
\label{PmetricCT2}
ds^{2}(M_6) = e^{-6\lambda}ds^{2} (T^2)  +  \frac{1-cy}{6}ds^{2} (S^2)
+ e^{-6\lambda }\sec^{2} \zeta d y^{2} + \frac{1}{9 } \cos^{2}
\zeta (D\psi +A)^{2},
\end{equation}
and, setting $L=1$ here for simplicity,
\begin{multline}\label{PmetricCT2G}
G_{(4)} = h_3 {vol}(S^2)vol(T^2)  +
\left[h_1 vol(S^2) +h_2 vol(T^2)\right]dy (D\psi +A)\\
+\left[\alpha_1 vol(S^2)+\alpha_2 vol(T^2)+\frac{1}{9}dy (D\psi+A)\right]F
+\frac{1}{3}dy (*_5 F) ,
\end{multline}
where $h_i=h_i(y)$ are the functions appearing in the undeformed flux \reef{4formypq} and
\bea
\alpha_1=\frac{(cy-1)y}{9} , \nn
\alpha_2=\frac{(cy-1)y}{3(a-y^2)} .
\eea

This can be dimensionally reduced and $T$-dualised (e.g. using the formula contained in
appendix C of \cite{gmmw3}) to obtain a Kaluza-Klein ansatz for the
$AdS_5\times Y^{p,q}$ solutions of type IIB.
We find:
\bea
ds^2&=&ds^2_5+\frac{1-cy}{6}ds^{2}(S^{2})+{e^{-6 \lambda }} \sec^{2} \zeta dy^{2}
+ \frac{1}{9} \cos^{2} \zeta (D \psi+A)^{2}\nn
&+&e^{6\lambda}[d\alpha-\frac{c}{6}A+\frac{-2y+y^2c+ac}{6(a-y^2)}(D\psi+A)]^2 , \nn
F_5&=&-4vol_5+(d\alpha-\frac{c}{6}A)[-\frac{2(1-cy)}{9}vol(S^2)dy(D\psi+A)+\frac{(cy-1)y}{9}vol(S^2)F\nn
&+&\frac{1}{9}dy(D\psi+A)F
+\frac{1}{3}dy (*_5F)]+\frac{1-cy}{18}vol(S^2)(*_5F)
+\frac{c}{18}dy(*_5)F(D\psi+A)\nn
&+&\frac{(cy-1)^2}{54}vol(S^2)F (D\psi+A) .
\eea
In this form  we clearly see that we have gauged shifts of the $\psi$ coordinate
as well as the $\alpha$ co-ordinate (when $c$ is non-zero). In order
to make the Reeb vector manifest, we can employ the coordinate change
$\alpha=-\beta/6-c\psi'/6$, $\psi=\psi'$ to get
\bea
ds^2&=&ds^2_5+\frac{1-cy}{6}ds^{2}(S^{2})+{e^{-6 \lambda }} \sec^{2} \zeta dy^{2}
+ \frac{1}{36} e^{6\lambda}\cos^{2} \zeta (d\beta+c\cos\theta d\phi )^{2}\nn
&+&\frac{1}{9}(d\psi'+{\cal A}+A)^2 , \nn
F_5&=&-4vol_5+\frac{1}{3}j(*_5F)-\frac{4}{3}vol_4(d\psi'+{\cal A} +A)+\frac{1}{9}j(d\psi'+{\cal A} +A)F ,
\eea
where $\theta$, $\phi$ parametrise $S^2$, and $j$, the K\"ahler-form on the locally K\"ahler-Einstein base space of the Sasaki-Einstein
space $Y^{p,q}$ orthogonal to the orbits of the Reeb vector $\partial_{\psi'}$, is given by
\be
j=\frac{1-cy}{6}vol(S^2)+\frac{1}{6}dy(d\beta+c\cos\theta d\phi) .
\ee
In addition, $vol_4$ is the volume form of the K\"ahler base, $vol_4=\frac{1}{2}j^2$, and
the potential ${\cal A}$, given by
${\cal A}=-\cos\theta d\phi +y(d\beta+c\cos\theta d\phi)$ satisfies
$d{\cal A}=6j$. This is precisely the ansatz discussed in \cite{Buchel:2006gb}. In particular
we see that we have gauged shifts of the coordinate $\psi'$ associated with the Reeb vector, which
is known to be dual to the $R$-symmetry.
This provides confirmation that the $M$-theory consistent truncation is indeed a reduction maintaining
the $R$-charge gauge field.

Let us now return to the consistent truncation for the most general solutions
and determine the correct normalisation of the gauge field.
The consistent truncation ansatz that we considered is invariant under the gauge transformations
\bea
\psi&\to\psi-\epsilon , \nn
A&\to A+d\epsilon .
\eea
To normalise the gauge field, following \cite{baryonspec}, we look for an object with known
$R$-charge. To do this we first think of our $AdS_5\times M_6$ solutions as special cases
of solutions of the form $M^{1,3}\times M_7$ where $M^{1,3}$ is four-dimensional
Minkowski space. As discussed in \cite{gmsw1} the relevant $M_7$ has a (local) $SU(3)$ structure
given by a vector $K'$, a three-form $\Omega'$ and a two-form $J'$
(see \cite{d7d6}). It was shown in \cite{G2spot} that such solutions
give rise to a superpotential of the form
\begin{equation}
\label{supersu3}
W \sim \int_{M_{7}}( G_{(4)}  + i K'   d J' )  \Omega',
\end{equation}
Since the superpotential has $R$-charge 2, we conclude that $\Omega'$ has
$R$-charge 2. Thus we just need to relate $\Omega'$
to the $SU(2)$ structure on $M_6$. In fact this was explicitly done in \cite{gmsw1}:
\begin{equation}
\Omega' = e^{-3 \lambda} \hat{\Omega}   (-\sin \zeta K_{1} - \cos
\zeta d r + i K_{2}) .
\end{equation}
The analysis of \cite{gmsw1} shows that the important $\psi$ dependence only appears
in the two-form $\hat\Omega$ and is of the form $\hat\Omega\sim e^{i\psi}$.
We thus conclude that $\psi/2$ is the appropriately normalised coordinate and
hence that $A/2$ is the appropriately normalised gauge field.

\end{document}